\documentclass[a4paper,11pt]{article} 
\pdfoutput=1 

\usepackage{jcappub} 
\usepackage[utf8]{inputenc} 

\usepackage{graphics}

\usepackage{graphicx} 

\usepackage{booktabs} 
\usepackage{array} 
\usepackage{paralist} 
\usepackage{verbatim} 
\usepackage{subfigure} 
\usepackage{amsmath}
\usepackage{amsfonts}
\usepackage{footnote}
\usepackage{stackengine}
\usepackage[toc,page]{appendix}

\title{\boldmath On preheating in $\alpha$-attractor models of inflation}

\author{Tomasz Krajewski,}
\author{Krzysztof Turzy\'nski}
\author{and Michał Wieczorek}

\affiliation{Institute of Theoretical Physics, Faculty of Physics, University of Warsaw, \\ Pasteura 5, 02-093 Warsaw, Poland}

\emailAdd{tomasz.krajewski@fuw.edu.pll}
\emailAdd{krzysztof.turzynski@fuw.edu.pl}
\emailAdd{michal.wieczorek@fuw.edu.pl}

\abstract{
We study (p)reheating in $\alpha$-attractor T-models of inflation, taking into account both scalar fields present in these models: the inflaton and the spectator. 
The two-field model has a negative field-space curvature which, at the end of inflation, may lead to geometrical destabilization of the spectator for small values of $\alpha\stackrel{<}{{}_\sim}10^{-3}$.
We perform the instability (Floquet) analysis of the linear dynamics and a fully non-linear lattice computations with our numerical code, which 
we specifically designed for a class of two-field models with non-canonical kinetic terms.
We find that the perturbations of the spectator field are much more unstable than the perturbations of the inflaton field,
so the dynamics of the early stages of preheating is dominated by the evolution of the spectator perturbations. 
As a result, the transition from the inflationary era to radiation domination era is practically instantaneous and much faster than previously found in an effective theory including only the inflaton field.
}

\keywords{inflation, preheating, lattice simulations, $\alpha$-attractors}

\arxivnumber{1801.01786}

\begin{document}
\maketitle
\flushbottom


\section{Introduction}

The concept of cosmological inflation has brought about a solution to many problems in big bang cosmology and, therefore, inflation has become a natural ingredient of the standard cosmological model 
(see e.g.\ \cite{VM} for a pedagogical introduction). 
However, inflation remains a very general theory and its relation with the Standard Model of particle physics is still unclear. 
In particular, the dynamics of the passage from the inflationary era to radiation-dominated (RD) era, called reheating, remains elusive. 
Already in pioneering works of Starobinsky the model of inflation came equipped with reheating
by gravitational particle creation in the regime of the weak narrow parametric resonance \cite{AStar1,AStar2} (see, {\it e.g.}, \cite{DeFelice:rev} for a review).
Once inflationary models employing a scalar field (dubbed the infllaton) coupled to general relativity became popular, models of particle production from
an inflaton condensate treated as a collection of unstable scalar particles were proposed \cite{OriRH1,OriRH2}. A radically different view, in which an effective classical force associated with the inflaton acting on quantum
fields leads to non-adiabatic excitations of scalar field fluctuations through parametric resonance,
was considered (for narrow parametric resonance)
in~\cite{BT}, but was later shown ineffective in the expanding Universe \cite{KLS}, while the same phenomenon driven by a broad parametric resonance \cite{KLS,PRH2,PRH3} remains a viable
candidate for the mechanism of reheating (see, {\it e.g.}, \cite{IR:rev} for a review). 

The multiplicity of approaches to reheating and the lack of observables that could 
at present distinguish between various possibilities have led many authors to consider reheating as an era of the evolution of the Universe
that is completely separate from inflation and to include the ignorance about that era into theoretical uncertainties in the predictions of inflationary models.
This is because the precise moment at which a given observed CMB mode had left the Hubble radius (which can be described as the number of e-folds between that instance and the end of inflation) 
is related to the subsequent evolution of the Universe, consisting of the (p)reheating period and the part of the RD era until that mode reenters the Hubble radius. The duration of the latter two phases depends on 
the evolution of the barotropic parameter $w=\langle p\rangle/\langle\rho\rangle$ during reheating and the reheating temperature, at which the $w$ approaches $1/3$ and the Universe enters the RD era. 
This in turn translates to the duration of the inflationary phase on which the basic inflationary predictions, the scalar spectral index $n_s$ and the tensor-to-scalar ratio $r$, depend. 
Therefore, the predictions of the inflationary models are often
expressed for a range of e-folds, typically $50-60$, which leads to $\mathcal{O}(10^{-2})$ uncertainty in the determination of $n_s$, comparable with observational uncertainty.
A more refined way of formulation inflationary predictions consists in considering all reasonable values of the (average) barotropic parameter $w$ and the reheating temperature \cite{EI}.

Certain inflationary models  come naturally equipped with a mechanism for efficient reheating.
A notable example is the class of $\alpha$-attractor models of inflation, called \mbox{T-models} \cite{CKL}, which have recently attracted a lot of interest because of several appealing features.
First, they are originally formulated in the context of supergravity, which gives them a solid theoretical motivation. 
Second, 
their predictions are naturally consistent with the Planck data  \cite{PC}. 
Last but not least, it has recently been realized that in a single-field effective theory of the inflaton field stemming from these models, at the end of inflation, 
the inflaton experiences self-resonance \cite{Aea} and its perturbations may become highly unstable;  once they dominate the Universe, the barotropic parameter can, for appropriate parameter choices, 
quickly approach $1/3$ \cite{AL,AL2}.
In this way, the radiation-dominated era begins very soon after the end of inflation,
which greatly reduces the theoretical uncertainty customarily attributed to the reheating era \cite{RH1,RH2,RH3,RH4,RH5,RH6,RH7,RH8} by, typically, an order of magnitude for $n_s$ \cite{AL}.

Even in the minimal supergravity construction, one should in principle consider both real degrees of freedom present in the scalar part of the chiral multiplet. 
Recently, a number of authors have studied the multi-field aspects of inflation in $\alpha$-attractor models \cite{alfamf0,alfamf1,alfamf2} and discussed the predictions
for the perturbations relevant for the CMB scales.
However, there are also interesting regions of the parameter space in which the scalar field that does not drive inflation (and that we shall from now on call the {\em spectator} field) can have important consequences for reheating. 
The spectator field is typically heavy when the CMB modes leave the Hubble radius, so its presence can be safely neglected for the calculation of the power spectrum of the curvature perturbations, but 
at the end of inflation the spectator field becomes transiently tachyonic and unstable, so its perturbations may eventually dominate the Universe.
This is possible because the noncanonical form of the kinetic part of the Lagrangian gives rise to geometrical destabilization \cite{RT}.
In the context of inflation, geometrical destabilization may end inflation prematurely \cite{GD1} or trigger a new phase of inflation \cite{GD2}. 
Also the dynamics of reheating can be affected by the instability of the spectator field and it is therefore interesting to study it in detail.

In this work, we present the analysis of preheating for \mbox{$\alpha$-attractor} T-models of inflation. 
We demonstrate that the spectator field may indeed become significant after the end of inflation.
We first show this semi-analytically, by performing a Floquet analysis for perturbations of both fields. 
Because the linearized equations of motion for the perturbations become unreliable when the instability kicks in, we
also present results of fully nonlinear lattice simulations of  preheating in these models. 
We 
find that the spectator field may have a very strong impact on the post-inflationary dynamics in these models. 
While this can make the reheating phase last even shorter than in the single-field description, at face value this effect does not directly affect the interpretation of the CMB data
in these models, because the uncertainty related to reheating is already removed thanks to the unstable dynamics of the inflaton field. However, it is interesting to understand which of the fields is the
principal driver of reheating, especially given the possibility of different coupling of these fields to matter fields.

Our work is organized as follows. In Chapter 2 we briefly present the $\alpha$-attractor \mbox{T-models} and analyze some of their features, which are crucial for an analysis of preheating. Chapter 3 is devoted to the Floquet analysis of the mode amplification during parametric resonance. In Chapter 4, we present the results of lattice simulations. We draw our conclusions in \mbox{Chapter 5}. A detailed description of the numerical procedure and results of additional simulations supplementing our main hypothesis are deferred to Appendices.

Throughout the paper we adopt natural units with $M_P=1$, unless indicated otherwise. 

\section{$\alpha$-attractor T-models of inflation}

\subsection{Presentation of the model}
\label{sec:alpha1}

We will 
consider
$\alpha$-attractor T-models of inflation
characterized by
the following superpotential
\begin{equation}\label{SuperPotential}
W_H=\sqrt{\alpha}\mu S\bigg(\frac{T-1}{T+1}\bigg)^n \, ,
\end{equation}
where $\mu$ is a constant parameter,
and by the K$\ddot{\text{a}}$hler potential
\begin{equation}\label{KahlerPotential}
K_H=-\frac{3\alpha}{2}\log\bigg(\frac{(T-\bar{T})^2}{4T\bar{T}}\bigg)+S\bar{S}
\end{equation}
with parameters $\alpha>0$ and $n>0$. As shown in \cite{CKLR}, the superfield $S$ can be stabilized during and after inflation and we can assume $S\equiv 0$. 
The scalar sector of the model can be then expressed in terms of two real scalar fields $\phi$ and $\chi$, which are related to the scalar component of the superfield $T$ by
\begin{equation}
\bigg|\frac{T-1}{T+1}\bigg|^2=\bigg(\frac{\cosh(\beta\phi)\cosh(\beta\chi) - 1}{\cosh(\beta\phi)\cosh(\beta\chi) + 1}\bigg)\quad\textrm{where}\quad\beta\equiv\sqrt{\frac{2}{3\alpha}}.
\end{equation}
This choice is justified by a particularly simple form of the field-space metric in the kinetic term of the scalar Lagrangian: 
\begin{equation}
\label{fsmetric}
\mathcal{L}=-\frac{1}{2}\Big(\partial_\mu\chi\partial^\mu\chi+\textrm{e}^{2b(\chi)}\partial_\mu\phi\partial^\mu\phi\Big)-V(\phi,\chi).
\end{equation}
where
$b(\chi)\equiv\log(\cosh(\beta\chi))$
and
the potential of the model reads 
\begin{equation}
\label{eq:potential}
V(\phi,\chi)=M^4\bigg(\frac{\cosh(\beta\phi)\cosh(\beta\chi) - 1}{\cosh(\beta\phi)\cosh(\beta\chi) + 1}\bigg)^n\Big(\cosh(\beta\chi)\Big)^{2/\beta^2} \,,
\end{equation}
with $M^4=\alpha\mu^2$.
With such a field-space metric,
the reparametrization invariant field-space curvarture
is negative
\begin{equation}\label{fcurvature}
\mathbb{R}=-2(b'^2+b'')=-2\beta^2=-\frac{4}{3\alpha}\,,
\end{equation}
which, in particular, means that one cannot canonically normalize both degrees of freedom.  Let us also note in passing that the supersymmetric origin of
the model considered herein allows applying the non-renormalization theorem for the superpotential, hence the form of the potential is
not affected by quantum corrections.

\begin{figure}
\centering
\includegraphics[width=0.47 \textwidth, height=0.35\textwidth]{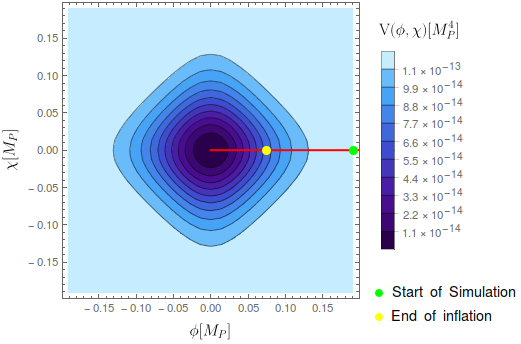}
\includegraphics[width=0.47 \textwidth, height=0.35\textwidth]{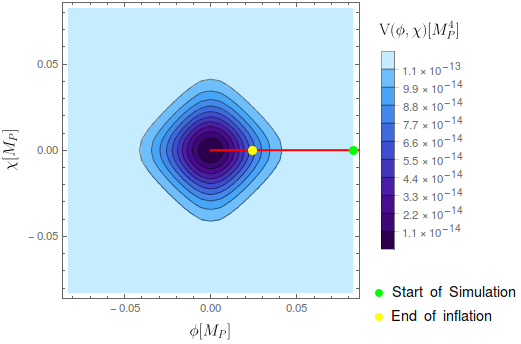} \\
\includegraphics[width=0.47 \textwidth, height=0.35\textwidth]{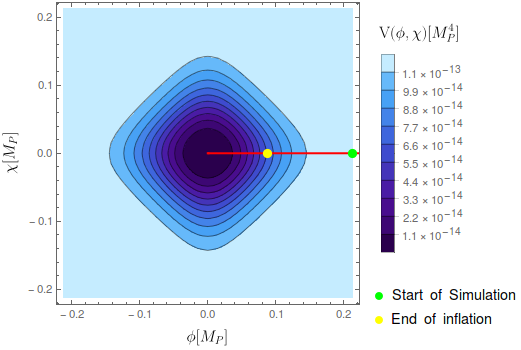}
\includegraphics[width=0.47 \textwidth, height=0.35\textwidth]{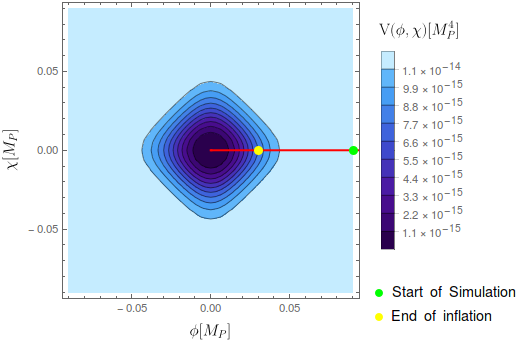}
\caption {\it Contour plots of the two-field potential (\ref{eq:potential}) for $n=1,\,\alpha=10^{-3}$ (upper left),  $n=1,\,\alpha=10^{-4}$ (upper right), $n=1.5,\,\alpha=10^{-3}$ (lower left),  $n=1.5,\,\alpha=10^{-4}$ (lower right). The red line represents the inflationary trajectory with the onset of the numerical simulations (described in Section \ref{LSimulations}) and the end of inflation marked as green and yellow dots, respectively.} 
\label{fig:pot}
\end{figure}

We shall present numerical results for four benchmark models, characterized in Table \ref{tab:benchmark}. These model have different values of the parameters $n$ and $\alpha$, which corresponds to
different shapes of the inflaton potential near the minimum and different strengths of the geometrical destabilization, respectively. The contour plots of the potentials are shown in
Figure~\ref{fig:pot}. Note that the potential (\ref{eq:potential}) has a plateau in the entire ($\phi$,$\chi$) plane away from the minimum.

\begin{table}
\centering
\begin{tabular}{|c|c|c|c|c|c|}
\hline
model & panel in figures & $n$ & $\alpha$ & $M$ & $k_\mathrm{max}$ \\
\hline
1 & upper left     & 1    & $10^{-3}$ & $5.96\times10^{-4}$ & $8.88\times10^{-5}$ \\
2 & upper right   & 1    & $10^{-4}$ & $3.37\times10^{-4}$ & $8.57\times10^{-5}$ \\
3 & lower left      & 1.5 & $10^{-3}$ & $5.97\times10^{-4}$ & $8.91\times10^{-5}$ \\
4 & lower right    & 1.5 & $10^{-4}$ & $3.38\times10^{-4}$ & $8.57\times10^{-5}$ \\
\hline
\end{tabular}
\caption{\it Description of the benchmark models used in the simulations. The last column shows the momentum space cutoff defined in Section \ref{sec:deslat}.}
\label{tab:benchmark}
\end{table}

\subsection{Inflationary trajectory and first-order perturbations}

As  shown in \cite{CKL}, 
the model presented in Section~\ref{sec:alpha1}
admits an inflating solution with the inflationary trajectory proceeding along $\chi=0$.
The model is therefore effectively described in terms of a canonically normalized inflaton field $\phi$ with a potential 
\begin{equation}
V(\phi, 0)=M^4\tanh^{2n}\bigg(\frac{\beta|\phi|}{2}\bigg).
\end{equation}
The model is consistent with Planck data  for a wide range of its parameters (see e.g. \cite{KLR}).
At this stage, $\phi$ is the inflaton and $\chi$ does not play any role in the evolution of the Universe.

This single-field description may cease to be adequate when the inflaton field accelerates and eventually leaves the slow-roll regime defined as $\epsilon\equiv-\dot{H}/H^2\ll 1$. 
A~negative value of field space curvature, can cause a `geometrical' destabilization of the perturbations of the field $\chi$ near the end of inflation \cite{RT},
which we identify with the moment at which $\epsilon$ reaches 1 for the first time. 
Therefore, in order to track the dynamics of the perturbations accurately, 
both fields should be taken into account.

Equations of motion for the perturbations in two-field models described by (\ref{fsmetric}) can be found e.g.\ in \cite{LLPT}, with no slow-roll approximation or any additional assumptions. 
The perturbed Friedmann-Robertson-Walker metric (in longitudinal gauge, with only scalar degrees of freedom included and constraints taken into account) reads
\begin{equation}
ds^2=-(1+2\Psi)dt^2+a^2(1-2\Psi)d\mathbf{x}^2 \, .
\end{equation}
From now on, we will assume $\chi=0$. The relevant equations of motion for the background quantities $H\equiv\dot{a}/a$ and $\phi(t)$ are
\begin{equation}
\label{eq:bkgd}
H^2=\frac{1}{3}\bigg[\frac{1}{2}\dot{\phi}^2+V(\phi,0)\bigg],\quad \ddot{\phi}+3H\dot{\phi}+V_{\phi}(\phi,0)=0\,,
\end{equation}
where $V_\phi$ stands for a derivative in the $\phi$ direction ($V_\phi\equiv \frac{\partial V}{\partial \phi}$).
Linear perturbations are described in terms of gauge invariant Mukhanov-Sasaki variables:
\begin{equation}
Q_\phi\equiv\delta\phi+\frac{\dot{\phi}}{H}\Psi\qquad \textrm{and}\qquad Q_\chi\equiv\delta\chi+\frac{\dot{\chi}}{H}\Psi,
\end{equation}
which obey the following equations of motion
\begin{equation}\label{RuchPert1}
\ddot{Q_\phi} + 3H\dot{Q_\phi} + \bigg(\frac{k^2}{a^2} + V_{\phi\phi}\bigg)Q_\phi=0\, ,
\end{equation} 
\begin{equation}\label{RuchPert2}
\ddot{Q_\chi} + 3H\dot{Q_\chi} + \bigg(\frac{k^2}{a^2} + m_\chi^2 \bigg)Q_\chi=0 \, ,
\end{equation}
where the effective mass of the spectator perturbation reads
\begin{equation}
\label{eq:mchi}
m_\chi^2 = V_{\chi\chi}+\frac{1}{2}\dot{\phi}^2\mathbb{R} \, .
\end{equation}
Writing eqs.\ (\ref{RuchPert1}), (\ref{RuchPert2}) and (\ref{eq:mchi}), we ignored all contributions suppressed by the Planck scale, as the energy scale of inflation is much smaller.
We also used the assumption $\chi=0$ which immediately implies that $V_{\chi}(\phi,0)=0$ and  $V_{\phi\chi}(\phi,0)=0$ for $V$ given by (\ref{eq:potential}).

\begin{figure}
\centering
\includegraphics[width=0.47 \textwidth, height=0.35\textwidth]{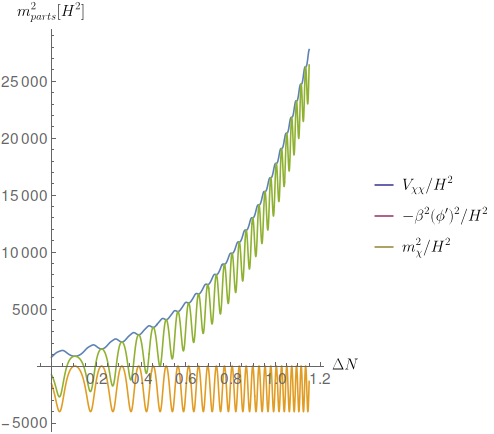}
\includegraphics[width=0.47 \textwidth, height=0.35\textwidth]{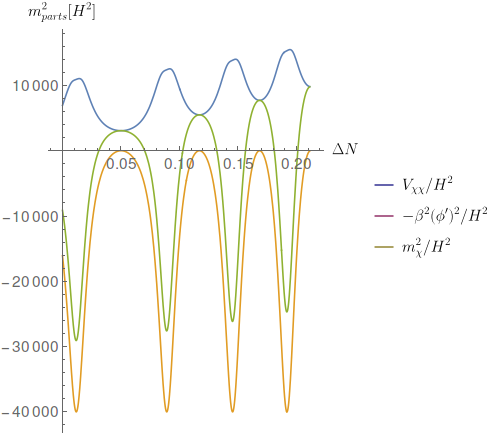} \\
\includegraphics[width=0.47 \textwidth, height=0.35\textwidth]{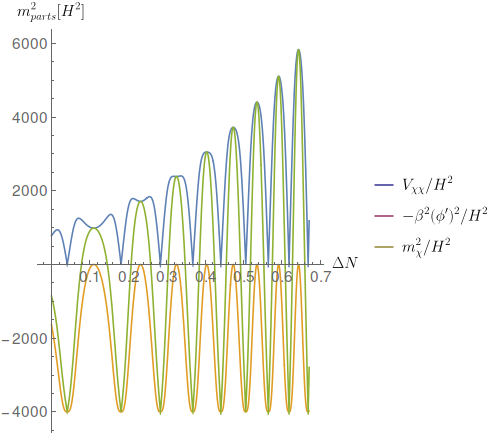}
\includegraphics[width=0.47 \textwidth, height=0.35\textwidth]{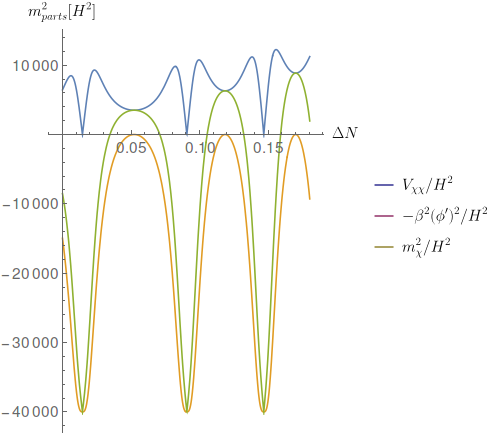}
\caption {\it Evolution of the mass of the spectator perturbation $\chi$ in Hubble units; from left to right for $n=1,\,\alpha=10^{-3}$,   $n=1,\,\alpha=10^{-4}$, $n=1.5,\,\alpha=10^{-3}$,  $n=1.5,\,\alpha=10^{-4}$; the evolution of the Hessian component and the geometrical component is also shown. $\Delta N$ is the
number of efolds that elapsed after the end of inflation.} 
\label{fig:chimass}
\end{figure}

Comparing eq. (\ref{fcurvature}) and (\ref{RuchPert2}) we can see that for large values of $\beta$, i.e.\ for small values of $\alpha$, 
we can expect the perturbation $Q_\chi$ to exhibit an {\it intermittent tachyonic instability} as $|\dot{\phi}|$ increases towards the end of inflation. 
In Fig.~\ref{fig:chimass}, we can see that this is indeed the case, as the mass of the spectator perturbation regularly assumes negative values
 after inflation. There are three main reasons behind a particular dependence of $m_\chi^2/H^2$ on time (measured by the number of
efolds $\Delta N$ after inflation). First, the Hubble parameter $H$ shows some wiggles, as the energy of the homogeneous component of the inflaton
decreases due to Hubble friction when the field rolls fast. Second, the product of $\beta=\sqrt{2/3\alpha}$ and the amplitude of the inflaton is much larger than
the Planck scale, which means that the argument of the hyperbolic cosine in (\ref{eq:potential}) is much larger than unity, so the potential cannot be reliably
expanded around $\phi=0$. Last but not least, the second term in (\ref{eq:mchi}), corresponding to the `geometrical' instability is always negative and oscillates with $\dot{\phi}^2$.

The time dependence of the mass of the spectator perturbations suggests that the {\it intermittent tachyonic instability of the spectator} may be an important factor for mode amplification and particle production, thereby competing with the parametric resonance of the inflaton perturbations, which was found to be effective for reheating \cite{AL,AL2}. 
We will analyze this instability quantitatively in Chapter~3, using Floquet theory. However, we would like to address first the issue of the initial conditions for perturbations, which will be important for our lattice simulations. 

\subsection{Initial conditions for perturbations}

One usually chooses the Bunch-Davies initial conditions for perturbations $Q_\phi$ and $Q_\chi$, appropriate for quantum fields in time-dependent, de Sitter background.  
This procedure is well known for fields with trivial field-space metric and is readily generalized for non-trivial cases. 
For two-field models with kinetic terms as in eq.~(\ref{fsmetric}),
the prescription for initial conditions of the perturbations can be found e.g.\ in \cite{LLPT}. 
This prescription becomes very simple, if one assumes
that $\chi=0$ and 
neglects
the contributions suppressed by the Planck scale. 
Defining $u_\phi\equiv aQ_\phi$ and $u_\chi\equiv aQ_\chi$, we can write the second-order action for the model as \cite{LRP}:
\begin{eqnarray}\nonumber
S_{2} & = & \frac{1}{2}\int d\tau d^3k\bigg[(u'_\phi)^2+(u'_\chi)^2-k^2u_\phi^2-k^2u_\chi^2\\
{} & {} & +a^2\bigg(2H^2-V_{\phi\phi}\bigg)u_\phi^2+a^2\bigg(2H^2-\bigg(V_{\chi\chi}+\frac{1}{2}\dot{\phi}^2\mathbb{R}\bigg)\bigg)u_\chi^2\bigg].
\end{eqnarray}
This action is a sum of two parts: one that depends only on $u_\phi$ and one that depends only on $u_\chi$. 
Each of these parts has the form of the action for a harmonic oscillator with time-dependent mass. 
Therefore, a usual single-field quantization procedure (see e.g.\ \cite{MW}) can be performed. In so-called adiabatic approximation, it provides the following initial conditions for perturbations:
\begin{equation}\label{IniPert1}
u_\phi(k,\tau_0) = \frac{1}{\sqrt{2\omega_{\phi,k}}}\exp^{-i\omega_{\phi,k}\tau_0},\qquad u'_\phi(k,\tau_0) = -i\sqrt{\frac{\omega_{\phi,k}}{2}}\exp^{-i\omega_{\phi,k}\tau_0}
\end{equation} 
and
\begin{equation}\label{IniPert2}
u_\chi(k,\tau_0) = \frac{1}{\sqrt{2\omega_{\chi,k}}}\exp^{-i\omega_{\chi,k}\tau_0},\qquad u'_\chi(k,\tau_0) = -i\sqrt{\frac{\omega_{\chi,k}}{2}}\exp^{-i\omega_{\chi,k}\tau_0},
\end{equation} 
where
\begin{equation}
\omega_{\phi,k}^2\equiv k^2+a^2\bigg(V_{\phi\phi}+2H^2\bigg)\quad \textrm{and}\quad \omega_{\chi,k}^2\equiv k^2+a^2\bigg(V_{\chi\chi}+2H^2+\frac{1}{2}\dot{\phi}^2\mathbb{R}\bigg).
\end{equation}
The formulae for energy density of perturbations per mode have the standard form
\begin{equation}\label{PertEnergy1}
E_\phi(k,\tau)=\frac{1}{2}\bigg(|u'_\phi(k,\tau)|^2+\omega_{\phi,k}^2|u_{\phi}(k,\tau)|^2\bigg)
\end{equation}
and
\begin{equation}\label{PertEnergy2}
E_\chi(k,\tau)=\frac{1}{2}\bigg(|u'_\chi(k,\tau)|^2+\omega_{\chi,k}^2|u_{\chi}(k,\tau)|^2\bigg).
\end{equation}
We shall use expressions \eqref{IniPert1} and \eqref{IniPert2} to set Gaussian initial conditions for perturbations in our lattice simulations.

\section{Floquet analysis of perturbations.}

From equations (\ref{PertEnergy1}) and (\ref{PertEnergy2}) it is clear that the amplitudes of fluctuations $u_\phi$ and $u_\chi$ (and hence $Q_\phi$ and $Q_\chi$) play a crucial role in the expression for energy density of perturbations. 
If these perturbations are unstable and their amplitudes grow sufficiently large, they can eventually dominate the energy density of the Universe, thereby affecting its equation of state.
On one hand, this can be treated as unavoidable uncertainty related to the reheating era. On the other hand, within each model the evolution of the perturbations can be in principle tracked numerically and the history of the Universe 
between the end of inflation and the onset of the radiation-dominated era can be reconstructed.

In this Section we shall discuss the growth of amplitudes of the perturbations semi-analytically,  using a Floquet analysis along the lines of Ref. \cite{FE}, deferring the full numerical analysis to Section~4.
We write eqs.~(\ref{RuchPert1}) and~(\ref{RuchPert2}) as two sets of two first order equations
\begin{equation}\label{Set1}
\left(\begin{array}{c}
\dot{Q}_{\phi,k} \\
\dot{\Pi}_{\phi,k} 
\end{array}\right)=
\left(\begin{array}{ccc}
0 & 1 \\
-\bigg(\frac{k^2}{a^2} + V_{\phi\phi}\bigg) & -3H
\end{array}\right)
\left(\begin{array}{c}
Q_{\phi,k} \\
\Pi_{\phi,k} 
\end{array}\right)
\end{equation}
and
\begin{equation}\label{Set2}
\left(\begin{array}{c}
\dot{Q}_{\chi,k} \\
\dot{\Pi}_{\chi,k} 
\end{array}\right)=
\left(\begin{array}{ccc}
0 & 1 \\
-\bigg(\frac{k^2}{a^2} + V_{\chi\chi}+\frac{1}{2}\dot{\phi}^2\mathbb{R}\bigg) & -3H
\end{array}\right)
\left(\begin{array}{c}
Q_{\chi,k} \\
\Pi_{\chi,k} 
\end{array}\right).
\end{equation}
After inflation the homogeneous field $\phi(t)$ begins to oscillate around the minimum of the potential. 
Because the timescale of these oscillation is typically much smaller that the timescale of the expansion of the Universe, 
$\phi(t)$ can be to a good approximation treated as a periodic function 
for time intervals spanning a few oscillations .
This implies that the matrices in eqs.~(\ref{Set1}) and (\ref{Set2}) are also periodic functions of time, as their time dependence comes primarily from their dependence on $\phi(t)$.
Therefore, by the Floquet Theorem, the fundamental matrices $\mathcal{O_{\phi,\chi}}(t,t_0)$ of solutions of the eqs.~\eqref{Set1} and~\eqref{Set2} can be written as
\begin{equation}\label{Floquet}
\mathcal{O_{\phi,\chi}}(t,t_0)=P_{\phi,\chi}(t,t_0)\exp{\Big[(t-t_0)\Lambda_{\phi,\chi}(t_0)\Big]},
\end{equation} 
where $P_{\phi,\chi}(t,t_0)$ are periodic matrices with the same period as matrices in eqs.~(\ref{Set1}) and (\ref{Set2}),
satisfying $P_{\phi,\chi}(t_0,t_0)=\mathbb{I}$.
Matrices $\Lambda_{\phi,\chi}(t_0)$ are constant (but $k$-dependent) matrices, whose eigenvalues $\mu_{\phi,\chi}^i$ are called Floquet exponents. 
A positive real part of a Floquet exponent indicates that the amplitude of the corresponding mode grows exponentially.
Therefore, a calculation of Floquet exponents for a range of modes and for different values of the amplitude of oscillating $\phi(t)$ may reveal which of these modes are unstable and, if both modes are unstable, which one grows faster.

\subsection{Floquet exponents for inflaton and spectator perturbations for $\alpha$-attractor \mbox{T-models}}
\label{sec:floo}

\begin{figure}
\centering
\includegraphics[width=0.47 \textwidth, height=0.35\textwidth]{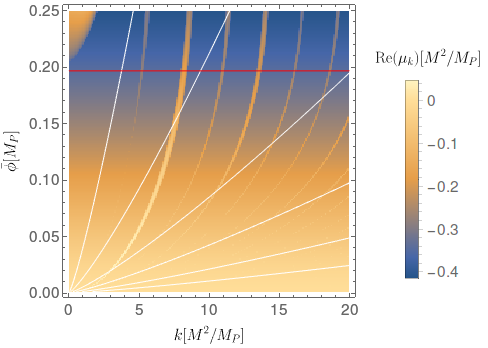}
\includegraphics[width=0.47\textwidth, height=0.35\textwidth]{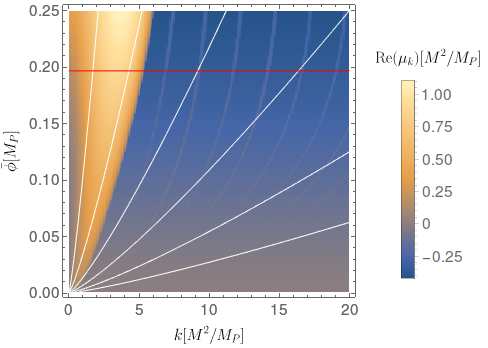}
\caption {\it Floquet exponents for the inflaton (left panel) and the spectator (right panel) perturbations with $n=3/2$ and $\alpha=10^{-2}$. } 
\label{FloquetExponents1}
\end{figure}

\begin{figure}
\centering
\includegraphics[width=0.47 \textwidth, height=0.35\textwidth]{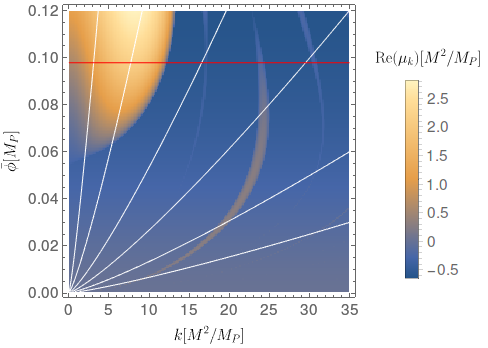}
\includegraphics[width=0.47\textwidth, height=0.35\textwidth]{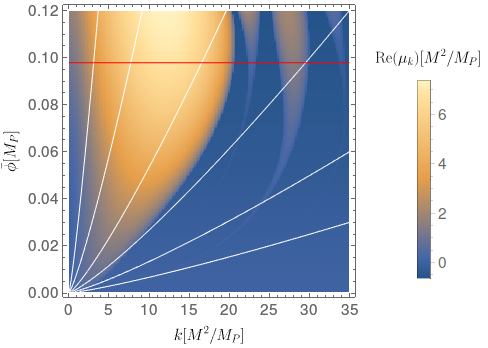}
\caption {\it Floquet exponents for the inflaton (left panel) and the spectator (right panel) perturbations with $n=3/2$ and $\alpha=10^{-3}$. } 
\label{FloquetExponents2}
\end{figure}

\begin{figure}
\centering
\includegraphics[width=0.47 \textwidth, height=0.35\textwidth]{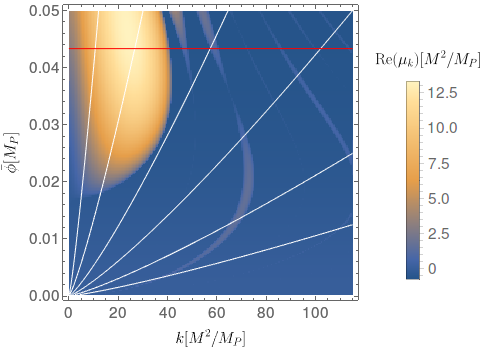}
\includegraphics[width=0.47\textwidth, height=0.35\textwidth]{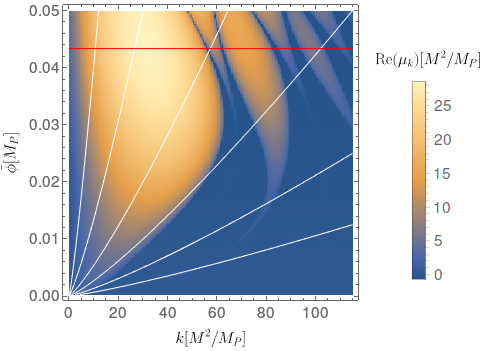}
\caption {\it Floquet exponents for the inflaton (left panel) and the spectator (right panel) perturbations with $n=3/2$ and $\alpha=10^{-4}$. } 
\label{FloquetExponents3}
\end{figure}

In Figures \ref{FloquetExponents1}-\ref{FloquetExponents3}, we present the Floquet exponents of inflaton and spectator perturbations for parameters $n=\frac{3}{2}$ and $\alpha=10^{-2}, 10^{-3}$ and $10^{-4}$.
Note that, unlike in Refs. \cite{AL,AL2}, in eqs.\ (\ref{Set1}) and (\ref{Set2}) we included the Hubble friction. In general, this leads to a decrease in the real parts of the resulting Floquet exponents by $\frac{3}{2}H$, 
so both can become negative.

We computed the Floquet exponents for a range of values of the amplitude $\bar{\phi}$ of the oscillating background field $\phi(t)$,
because if we follow the evolution of the Universe for many efolds, the Hubble friction leads to a slow
decrease of this amplitude, which is described by the relation $\bar{\phi}\propto a^{-3/(n+1)}$.
Similarly, in the expanding Universe the effective wave number decreases and satisfies $k_{\mathrm{eff}}=k/a$. Therefore, an initial condition at the end of inflation corresponds to a particular value of the amplitude of the homogeneous inflaton field, which in the plots is indicated with a red line. As the Universe expands, a given mode corresponds to a certain path on the $(k_\mathrm{eff},\bar{\phi})$ plane. 
Therefore, to describe the growth of the particular modes during their evolution, we need to
compute Floquet exponents for different amplitudes and wave numbers. In Figures \ref{FloquetExponents1}-\ref{FloquetExponents3}, the $(k_{\mathrm{eff}},\bar{\phi})$-paths for a few different modes are drawn as white curves.

For a discussion of the Floquet exponents, it is convenient to
introduce explicitly the reduced Planck mass $M_P$ to keep track of mass dimensions of different quantities.
Values of Floquet exponents are then given in units $M^2/M_{P}$.  
These units are natural for Floquet exponents for two reasons. 
First, the values of Floquet exponents are \mbox{$M$-independent} in such units. 
Second, the Hubble rate is of order of $M^2/M_{P}$ at the end of inflation. 
Therefore, values of Floquet exponents describe naturally the rate of the exponential growth of the amplitude of a given mode: 
with $\tilde{\mu}_k\equiv\frac{\mu_k}{M^2/M_P}$, 
the amplitude grows roughly by $\sim e^{\tilde{\mu}_k}$ during one Hubble time.

In our computation of Floquet exponents we have taken into account the expansion of the Universe. 
We used the full equation of motion (\ref{eq:bkgd}) for the background quantities and eqs.~(\ref{Set1}) and (\ref{Set2}) for the perturbations.
This, in particular, implies that $\phi(t)$ is an almost periodic function modulated by a slowly decreasing envelope. We calculated the Floquet exponents in a standard way, comparing the values of $Q_{\phi,k}$ and $\Pi_{\phi,k}$
at two subsequent maxima of $\phi(t)$; the same calculation was applied for $Q_{\chi,k}$ and $\Pi_{\chi,k}$.
This shifts the obtained values of Floquet exponents by $-\frac{3}{2}H$ with respect to the computation with no expansion included. 
Taking this into account, the Floquet exponents for the inflaton shown in Figures~\ref{FloquetExponents1}-\ref{FloquetExponents3} are consistent with the results presented in \cite{AL,AL2} and with a more
recent analysis \cite{Iarygina:2018kee}.

Let us summarize briefly the results shown in Figs.~\ref{FloquetExponents1}-\ref{FloquetExponents3}. 
For all cases, the maximal Floquet exponents for the spectator are much larger than those for the inflaton. 
Also, the regions on the $(\bar{\phi},k)$ plane in which the Floquet exponents exceed the Hubble parameter are much larger for the spectator than for the inflaton. This is because the
spectator pertubations, unlike the inflaton perturbations, have a `geometrical'  tachyonic instability that sets in for the spectator field shortly before the end of inflation \cite{RT}.
Combining these two facts, we conclude that there is an important difference between the single- and two-field descriptions of preheating. Had the spectator field been absent,
the inflaton perturbations would have been excited by a parametric resonance, as described in \cite{AL,AL2}. However, in the two-field calculation, the spectator is subject
to an intermittent tachyonic instability, which may be so strong that it rapidly drives the perturbation beyond the linear regime. While our Floquet analysis is useful to identify
certain aspects of the dynamics of the two-field system in the linear regime at the early stage of the instability, as 
the perturbations grow -- and interact -- beyond the linear level, we need to study the evolution of the perturbations resorting to fully non-linear lattice simulations.


\section{Lattice Simulations}\label{LSimulations}

\subsection{Description of simulations}
\label{sec:deslat}

There are many computer codes written to simulate preheating after inflation on the lattice (see e.g~\cite{FE} for a recent review). 
However, most of them can be applied only to models with canonical kinetic term in the Lagrangian (e.g.~\cite{S,LE,DF}). 
One exception is GABE \cite{GABE}, designed to make simulations for models with non-canonical kinetic terms. However, this code is based on the Runge-Kutta method, which is not symplectic
and turned out unsuitable for our purposes.
Using a non-symplectic method
significantly decreases accuracy of the long-time simulations and spoils energy conservation. 
Fortunately, the $\alpha$-attractor T-models field space metric (\ref{fsmetric}) is very particular in the sense that one can construct the explicit symplectic method in this case. 
Its description is provided in \mbox{Appendix \ref{Numerics}.}
We performed the simulations using a code written by ourselves which is based on this method.
At this point, our goal was to perform an exploratory analysis to catch a glimpse of the behavior of the spectator perturbations. 
Therefore, we used cubic lattices of a quite modest size:
throughout the paper, we report on results obtained with $N_\mathrm{lattice}=128$, unless indicated otherwise. We also checked that results of our
computations are stable against reducing the lattice size to $N_\mathrm{lattice}=64$.
We performed the simulations for $\alpha=10^{-3}$ and $\alpha=10^{-4}$, where we expected strong instability of the spectator, based on the results of the Floquet analysis presented in Section \ref{sec:floo}. 
We used the momentum space cutoff $k_\mathrm{max}=250M^2/M_P$ and $k_\mathrm{max}=750M^2/M_P$ for $\alpha=10^{-3}$ and $\alpha=10^{-4}$ respectively, which we found to be a good trade-off between granularity in Floquet instability regions and factoring in higher frequency modes. Our physical results are of course insensitive to the choice of the cutoff, provided that it is large enough, as we demonstrate in Appendix \ref{app:b12}. The implied lattice spacing is $h=\sqrt{3}\pi/k_\mathrm{max}$.

In our model the definition of the barotropic parameter $w$ can be written in terms of fields as
\begin{equation}
w\equiv\frac{\langle p\rangle}{\langle\rho\rangle}=\frac{\bigg\langle(\frac{\textrm{e}^{2b(\chi)}\dot{\phi}^2+\dot{\chi}^2)}{2}-\frac{(\textrm{e}^{2b(\chi)}(\nabla\phi)^2+(\nabla\chi)^2)}{6a^2}-V(\phi,\chi)\bigg\rangle}{\bigg\langle\frac{(\textrm{e}^{2b(\chi)}\dot{\phi}^2+\dot{\chi}^2)}{2}+\frac{(\textrm{e}^{2b(\chi)}(\nabla\phi)^2+(\nabla\chi)^2)}{2a^2}+V(\phi,\chi)\bigg\rangle},
\end{equation}
where the brackets $\langle\rangle$ denote the (optional) time average over a few oscillations of the homogeneous inflaton field $\phi(t)$. 
If the potential for the inflaton behaves as $V\sim |\phi|^{2n}$ around the minimum and the Universe is dominated by the homogeneous inflaton field $\phi$, the barotropic parameter satisfies $w=\frac{n-1}{n+1}$ \cite{VM}, which for our
benchmark models would give $w=0.2$ ($w=0$) for $n=1.5$ ($n=1$). However, it has been shown in Ref.\ \cite{AL} that in the single-field $\alpha$-attractor \mbox{T-model}, the growth of the inflaton perturbations due to self-resonance \cite{Aea} can be
so large that for $n=1.5$ these perturbations dominate the Universe which expands according to the equation of state with $w=1/3$, i.e.~it evolves as a radiation dominated (for $n=1$ oscillons are formed and the effective equation of state with $w=0$ does not change). 
Since we found that the growth of the spectator perturbations estimated from the Floquet analysis is faster that the growth of the inflaton perturbations, we may hypothesize that spectator perturbations are the 
main factor that makes the Universe approach the radiation-like state with $w=1/3$. Such a hypothesis can only be tested by means of numerical simulations.

\subsection{Results}
\label{sec:res}

\begin{figure}
\centering
\includegraphics[width=0.495 \textwidth, height=0.29\textwidth]{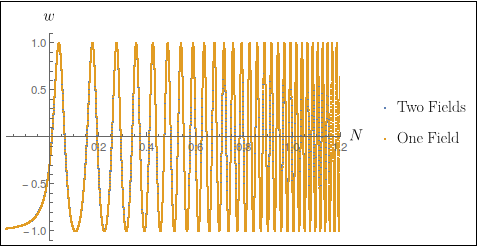}
\includegraphics[width=0.495 \textwidth, height=0.29\textwidth]{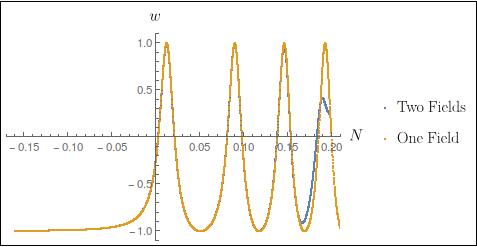} \\
\includegraphics[width=0.495 \textwidth, height=0.29\textwidth]{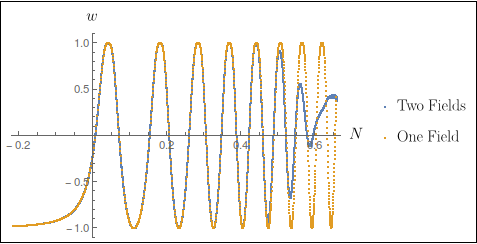}
\includegraphics[width=0.495 \textwidth, height=0.29\textwidth]{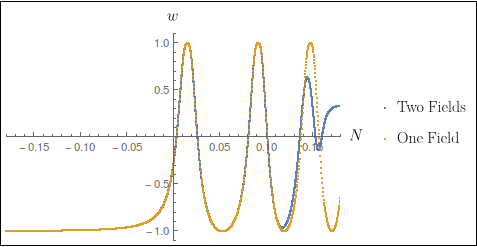}
\caption{\it The evolution of the barotropic parameter $w$ (not averaged) for 
 $n=1,\,\alpha=10^{-3}$ (upper left),  $n=1,\,\alpha=10^{-4}$ (upper right), $n=1.5,\,\alpha=10^{-3}$ (lower left),  $n=1.5,\,\alpha=10^{-4}$ (lower right). 
$N$ is the number of e-folds after the end of inflation. Blue points mark the results for a full two-field calculation, while orange points give predictions of the one-field model with $\chi=0$. }
\label{wPlot} 
\end{figure}

Lattice simulations of preheating for a single field $\alpha$-attractor T-models (i.e.\ without the perturbations of $\chi$) have been already performed by the authors of Ref.\ \cite{AL}. As one of the tests of our code, we repeated these simulations and our results are in agreement with those shown in~\cite{AL}. We show in Figure~\ref{wPlot} the evolution of barotropic parameter $w$ calculated within the four benchmark models described in Section~\ref{sec:alpha1}. 


\begin{figure}
\centering
\includegraphics[width=0.495 \textwidth, height=0.2\textwidth]{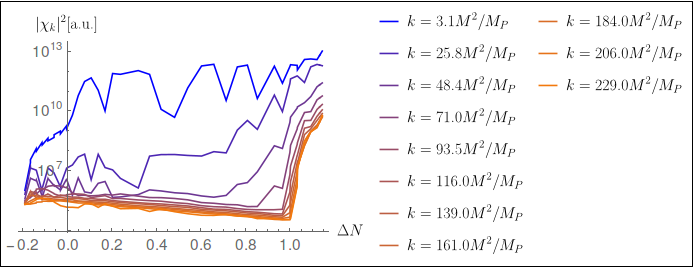} 
\includegraphics[width=0.495\textwidth, height=0.2\textwidth]{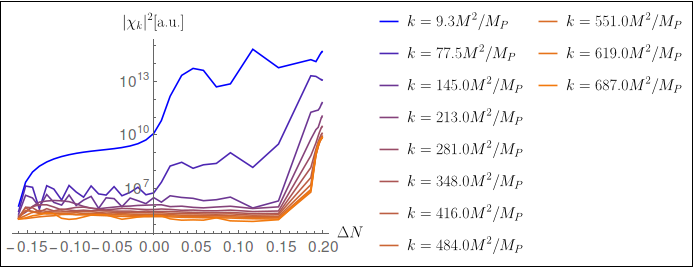} \\
\includegraphics[width=0.495 \textwidth, height=0.2\textwidth]{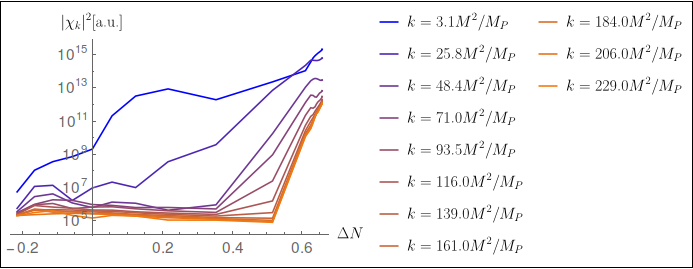} 
\includegraphics[width=0.495\textwidth, height=0.2\textwidth]{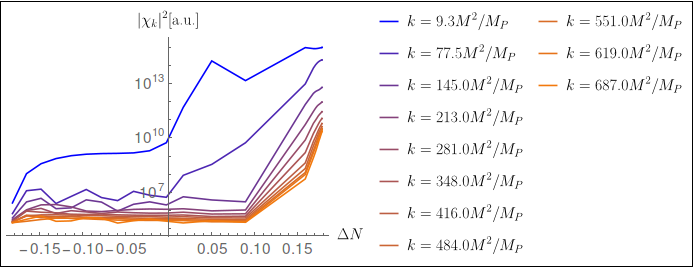} \\
\caption{\it The time evolution of the spectator perturbations for 
 $n=1,\alpha=10^{-3}$ (upper left),  $n=1,\alpha=10^{-4}$ (upper right), $n=1.5,\alpha=10^{-3}$ (lower left),  $n=1.5,\alpha=10^{-4}$ (lower right)
for different number of wavenumbers~$k$.} 
\label{TimeEvolution}
\end{figure}

\begin{figure}
\centering
\includegraphics[width=0.495 \textwidth, height=0.2\textwidth]{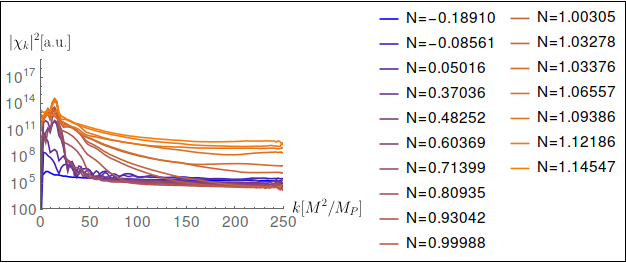} 
\includegraphics[width=0.495\textwidth, height=0.2\textwidth]{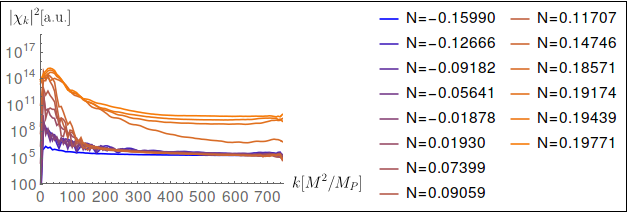} \\
\includegraphics[width=0.495 \textwidth, height=0.2\textwidth]{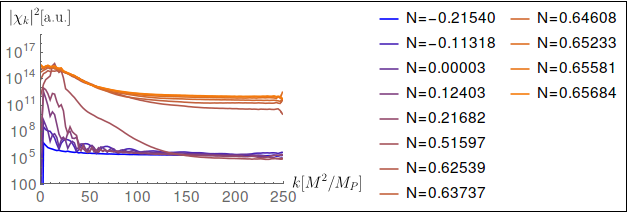} 
\includegraphics[width=0.495\textwidth, height=0.2\textwidth]{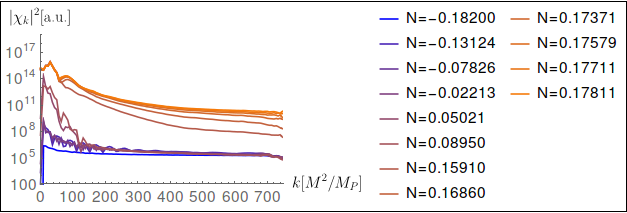} \\
\caption{\it The power spectrum of the spectator perturbations for 
 $n=1,\alpha=10^{-3}$ (upper left),  $n=1,\alpha=10^{-4}$ (upper right), $n=1.5,\alpha=10^{-3}$ (lower left),  $n=1.5,\alpha=10^{-4}$ (lower right)
for different number of e-folds $N$ after the end of inflation.} 
\label{PowerSpectrum}
\end{figure}

If perturbations of the inflaton become so important in single-field simulations, it is quite reasonable to expect, that the more strongly amplified perturbations of the spectator may be the main force that drives the Universe towards a radiation-like state. Indeed, in two-field simulations, the growth of spectator perturbations for the values of parameters that gave effective reheating in the single-field case is so strong that 
our variable-step simulations typically stall within an efold after the end of inflation, while our fixed step simulations report unacceptably large errors. 
However, 
before that happens
our determination of the barotropic parameter $w$ is physically relevant and we are allowed to conclude that tachyonic instability of the spectator causes a very fast growth of the barotropic parameter just after the end of inflation, with $w$ approaching 1/3 in all benchmark models except for $n=1,\,\alpha=10^{-3}$ benchmark model\footnote{However, results of calculations on a smaller lattice $N_\mathrm{lattice}=64$ presented in Appendix \ref{app:b0} strongly suggest that $w$ approaches 1/3 already at $N=1.2$ efolds after inflation in this benchmark model}. In this sense, in these models reheating may be completed much faster than it follows from single-field simulations. We therefore find that the presence of the spectator leads to practically immediate reheating both for $n=1$ and $n=1.5$.


\begin{figure}
\centering
\includegraphics[width=0.495 \textwidth, height=0.2\textwidth]{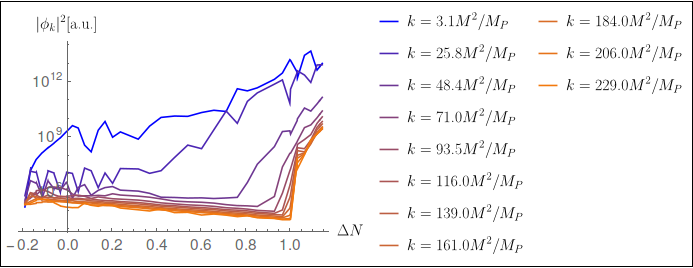} 
\includegraphics[width=0.495\textwidth, height=0.2\textwidth]{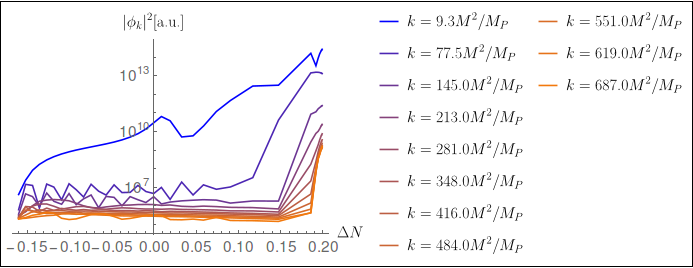} \\
\includegraphics[width=0.495 \textwidth, height=0.2\textwidth]{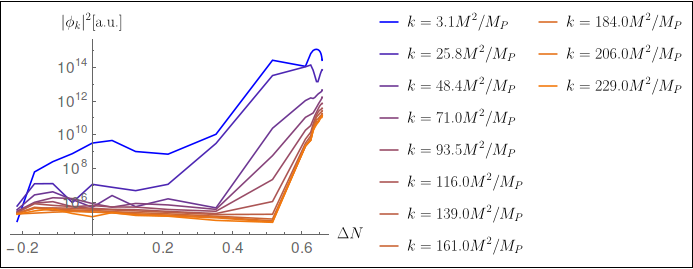} 
\includegraphics[width=0.495\textwidth, height=0.2\textwidth]{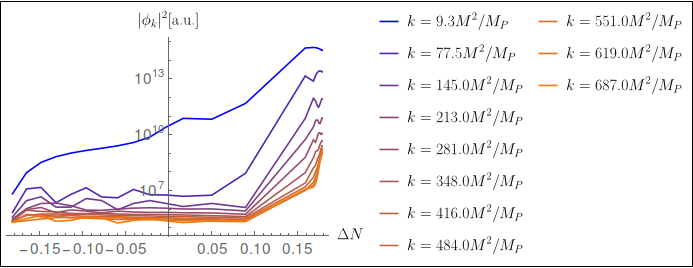} \\
\caption{\it The time evolution of the inflaton perturbations for 
 $n=1,\alpha=10^{-3}$ (upper left),  $n=1,\alpha=10^{-4}$ (upper right), $n=1.5,\alpha=10^{-3}$ (lower left),  $n=1.5,\alpha=10^{-4}$ (lower right)
for different number of wavenumbers~$k$.} 
\label{TimeEvolution2}
\end{figure}

\begin{figure}
\centering
\includegraphics[width=0.495 \textwidth, height=0.2\textwidth]{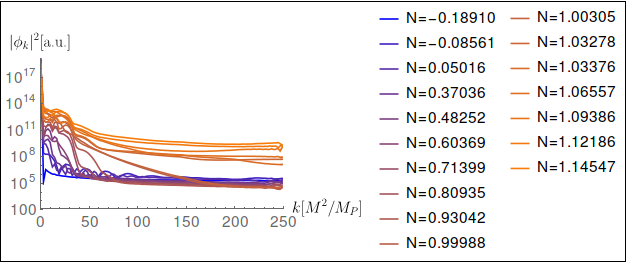} 
\includegraphics[width=0.495\textwidth, height=0.2\textwidth]{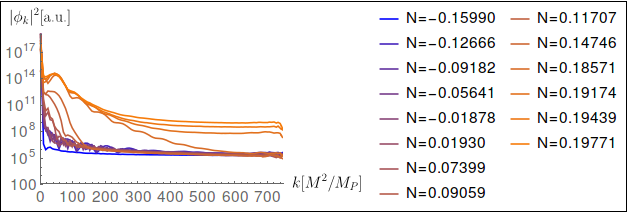} \\
\includegraphics[width=0.495 \textwidth, height=0.2\textwidth]{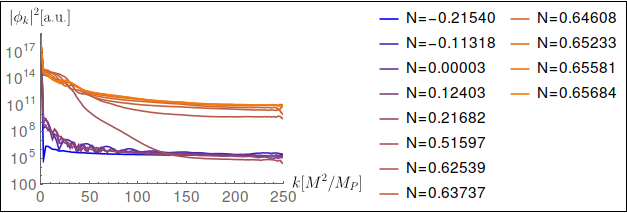} 
\includegraphics[width=0.495\textwidth, height=0.2\textwidth]{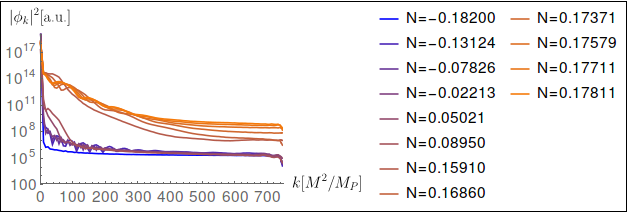} \\
\caption{\it The power spectrum of the inflaton perturbations for 
 $n=1,\,\alpha=10^{-3}$ (upper left),  $n=1,\,\alpha=10^{-4}$ (upper right), $n=1.5,\,\alpha=10^{-3}$ (lower left),  $n=1.5,\,\alpha=10^{-4}$ (lower right)
for different number of e-folds $N$ after the end of inflation.} 
\label{PowerSpectrum2}
\end{figure}

The growth of spectator perturbations can be investigated further with Fourier analysis
In Figure~\ref{TimeEvolution}, we show the time evolution of a few Fourier modes for our four benchmark models.
In Figure \ref{PowerSpectrum}, we show the plots of the power spectrum of the spectator field for different moments after the end of inflation for our four benchmark models. 
At first, only the modes with wavenumbers $k$ smaller than the maximal absolute value of the negative spectator mass are excited very strongly. 
A closer look also reveals
subsequent rescattering resulting in a `ringing' pattern of the amplified modes.
After initial growth these modes backreact and cause the growth of modes with larger wavenumber~$k$. The obtained initial growth of small-$k$ modes is consistent with the approximation obtained from the linear Floquet analysis \footnote{For a back-of-the-envelope estimate, we can consider as an example 
values of the spectator power spectrum for $n=1.5, \alpha=10^{-4}$ and $k_{\mathrm{eff}}=60M^2/M_{P}$ at two different moments: $N=0.01$, $N=0.09$ efolds after inflation. Then from the Floquet analysis, we obtain:
\begin{equation}
\frac{|\chi_{N=0.09}|^2}{|\chi_{N=0.01}|^2}\approx\Bigg(\exp\bigg(\langle\mu_{k_{\mathrm{eff}}}\rangle\frac{\Delta N}{H}\bigg)\Bigg)^2\approx\exp\bigg(25\frac{M^2}{M_{P}}\sqrt{3}\frac{M_{P}}{M^2}\cdot 0.08\cdot 2\bigg)\approx 10^3
\end{equation} 
This value is in agreement with results shown on the right panel in Figure \ref{PowerSpectrum}.}.
The fact that the growth of these modes is faster for $\alpha=10^{-4}$ than for $\alpha=10^{-3}$ can be easily understood, since the smaller $\alpha$ is, the stronger tachyonic instability the spectator exhibits.



A gradual growth of higher frequency modes cannot be predicted by the Floquet theory, since it is a purely nonlinear effect. 
However, this effect plays a very important role in reheating, since it leads to the fragmentation of fields. On the timescales considered here, this crucially depends on the evolution of the field $\chi$ that
drives the instability; as the two fields $\phi$ and $\chi$ are tightly coupled through the non-canonical kinetic term, higher frequency modes of $\phi$ quickly follow those of $\chi$, which can be seen comparing Figures \ref{PowerSpectrum} and \ref{PowerSpectrum2}.

\begin{figure}
\centering
\includegraphics[width=0.495 \textwidth, height=0.29\textwidth]{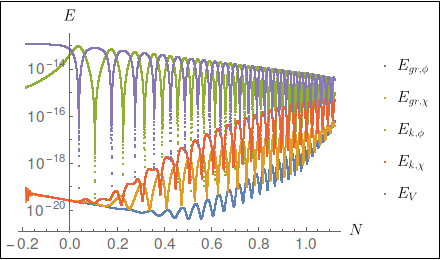}
\includegraphics[width=0.495 \textwidth, height=0.29\textwidth]{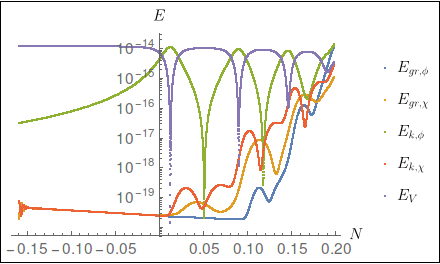} \\
\includegraphics[width=0.495 \textwidth, height=0.29\textwidth]{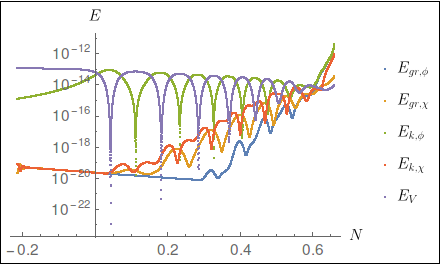}
\includegraphics[width=0.495 \textwidth, height=0.29\textwidth]{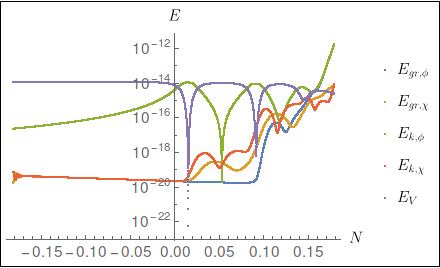}
\caption{\it The evolution of the components of the energy density for 
 $n=1,\,\alpha=10^{-3}$ (upper left),  $n=1,\,\alpha=10^{-4}$ (upper right), $n=1.5,\,\alpha=10^{-3}$ (lower left),  $n=1.5,\,\alpha=10^{-4}$ (lower right). 
$N$ is the number of e-folds after the end of inflation. }
\label{fig:en} 
\end{figure}

For completeness, we also show in Figure \ref{fig:en}, the evolution of different components of the energy density for our four benchmark models. Except for $n=1,\,\alpha=10^{-3}$, for which the simulations end prematurely, we find that kinetic and gradient energies of the fields start domination very quickly. That the contributions coming from the perturbations of $\phi$ dominates over those of $\chi$ follows from the
factor of $\cosh^2(\beta\chi)$ multiplying the kinetic term of $\phi$. 

In the benchmark models studied in this paper, a strong instability of the spectator perturbations results from a sufficiently small value of $\alpha$ and a large negative `geometrical' contribution to the effective mass
of the spectator perturbations. It is therefore interesting to check for which values of $\alpha$
this instability immediately leads to a radiation-dominated-like era, i.e.~$w\to1/3$. To this end, we performed a series of simulations on smaller lattices with $N_\mathrm{lattice}=64$ in which we kept $n=1.5$ fixed and we
varied $\alpha$. We found that for $\alpha\stackrel{>}{{}_\sim}2\times 10^{-3}$ there is no immediate radiation-dominated-like era and the barotropic parameter $w$ approaches $0.2$, as in the single-field case (before self-resonance). 


\section{Conclusions and outlook}

In this paper we examined the issue of preheating for $\alpha$-attractor T-models of inflation. 
%
We have shown that during preheating the spectator field may play an important role, driving the exponential growth of the energy density and pressure perturbations. 
Results of our numerical simulations indicate that the {\it intermittent tachyonic instability} of the spectator is the main factor driving the mode amplification. As a result,
for $\alpha\stackrel{<}{{}_\sim}10^{-3}$, the Universe enters a radiation-dominated-like phase, characterized by the equation of state $p\sim\rho/3$, which suggests that the reheating process lasts shorter
and is more generic than previously thought.

For very small values of the parameter $\alpha$, the growth of perturbations is so strong that it is very hard to tract them numerically,
therefore, we treat our result as a first step in the exploration of the dynamics of reheating in $\alpha$-attractor T-models of inflation. 
In particular, it remains to be seen whether the inflaton condensate can decay completely into unstable perturbations, thus completing
the reheating process.
We also note that
obtained overdensities are so huge that 
they can possibly lead to primordial black holes formation, which may lead to lower bounds on $\alpha$. 
Albeit very interesting, these issues lie beyond the scope of the present note and we shall address them in future analyses,
which would require employing larger lattice sizes and using more powerful computational facilities.

\subsubsection*{Acknowledgements}

T.K. is supported by grants No.\ DEC-2012/04/A/ST2/00099 and 2016/23/N/ST2/0311 from the National Science Centre (Poland).
M.W. and K.T. are supported by grant No.\ 2014/14/E/ST9/00152 from the National Science Centre (Poland).

\begin{appendix}

\section{The symplectic numerical method to simulate preheating in the $\alpha$-attractor T-model of inflation}\label{Numerics}

The method presented here is a modification of the method presented in description of PyCOOL \cite{S} -- the lattice code for simulating preheating for scalar fields inflationary models. PyCOOL can be used only for models with canonical kinetic term in their Lagrangian. Therefore, for $\alpha$-attractor T-models, we wrote a code based on our modified method.

\subsection{Action and Hamiltonian}

Our goal is to solve numerically equations which come from the action
\begin{equation}\label{action2}
S=\int d^4x\sqrt{-g}\bigg[\frac{1}{2}\mathcal{R}-\frac{1}{2}e^{2b(\chi)}(\partial_\mu\phi)(\partial^\mu\phi)-\frac{1}{2}(\partial_\mu\chi)(\partial^\mu\chi)-V(\phi,\chi)\bigg]
\end{equation}
We assume here that the spacetime is spatially homogeneous, isotropic and flat, i.e.~we have
\begin{equation}
ds^2=a^2(-d\tau^2+d\mathbf{x}^2),
\end{equation}
which implies
\begin{equation}
\sqrt{-g}=a^4\qquad \textrm{and}\qquad \mathcal{R}=6\frac{a''}{a^3},
\end{equation}
where the prime denotes the derivative with respect to the conformal time $\tau$.
After discretization in space the action \eqref{action2} can be written as:
\begin{eqnarray}\nonumber
S & = & (dx)^3\int\mathcal{L}d\tau =\\ 
{} & = & (dx)^3\int\bigg[-3a'^2V_L+\sum_{\vec{x}}\frac{a^2}{2}\Bigg(e^{2b(\chi_{\vec{x}})}\bigg((\phi'_{\vec{x}})^2-\frac{G(\phi,\vec{x})}{(dx)^2}\bigg)+\\ \nonumber
{} & + & \bigg((\chi'_{\vec{x}})^2-\frac{G(\chi,\vec{x})}{(dx)^2}\bigg)-a^2V(\phi_{\vec{x}},\chi_{\vec{x}})\Bigg)\bigg]d\tau,
\end{eqnarray} 
where $(dx)^3V_L$ equals the volume  of the periodic lattice and 
\begin{equation}
G(Y,\vec{x})=\frac{1}{2}\sum_{x_1-1}^{x_1+1}\sum_{x_2-1}^{x_2+1}\sum_{x_3-1}^{x_3+1}c_{d(\alpha)}(Y_\alpha-Y_0)^2
\end{equation}
is the second order discretization of the squared spatial gradient operator
\begin{equation}
(\nabla Y)^2(\vec{x})\simeq\frac{G(Y,\vec{x})}{(dx)^2}
\end{equation}
with $c_1=1$ and $c_0=-6$ (see \cite{DF}).\\
After the Legendre transformation we obtain the following Hamiltonian density:
\begin{equation}
\mathcal{H}=-\frac{p_a^2}{12V_L}+\sum_{\vec{x}}a^4\bigg(\frac{\pi^2_{\phi,\vec{x}}}{2a^6e^{2b(\chi_{\vec{x}})}}+
\frac{\pi^2_{\chi,\vec{x}}}{2a^6}+e^{2b(\chi_{\vec{x}})}\frac{G(\phi,\vec{x})}{2(dx)^2a^2}+\frac{G(\chi,\vec{x})}{2(dx)^2a^2}+V(\phi_{\vec{x}},\chi_{\vec{x}})\bigg),
\end{equation} 
where canonical momenta are defined by the formulae
\begin{equation}
p_a\equiv\frac{\partial\mathcal{L}}{\partial a'}=-6a'V_L,\quad \pi_{\phi,\vec{x}}\equiv \frac{\partial\mathcal{L}}{\partial \phi'}=a^2\textrm{e}^{2b(\chi_{\vec{x}})}\phi'_{\vec{x}}\quad\textrm{and}\quad \pi_{\chi,\vec{x}}\equiv \frac{\partial\mathcal{L}}{\partial \chi'}=a^2\chi'_{\vec{x}}.
\end{equation}

\subsection{Discretization scheme}
The time upgrade scheme is based on the fact that we can divide this Hamiltonian into four parts
\begin{equation}
\mathcal{H}=\mathcal{H}_1+\mathcal{H}_2+\mathcal{H}_3+\mathcal{H}_4
\end{equation}
in such the way that none of the parts depends on both the field and its canonical momentum. An example of such division is
\begin{equation}
\mathcal{H}_1\equiv -\frac{p_a^2}{12V_L},
\end{equation}
\begin{equation}
\mathcal{H}_2\equiv \sum_{\vec{x}}a^4\bigg(\frac{\pi^2_{\phi,\vec{x}}}{2a^6e^{2b(\chi_{\vec{x}})}}\bigg),
\end{equation}
\begin{equation}
\mathcal{H}_3\equiv\sum_{\vec{x}}a^4\bigg(\frac{\pi^2_{\chi,\vec{x}}}{2a^6}\bigg)
\end{equation}
and
\begin{equation}
\mathcal{H}_4\equiv \sum_{\vec{x}}a^4\bigg(e^{2b(\chi_{\vec{x}})}\frac{G(\phi,\vec{x})}{2(dx)^2a^2}+\frac{G(\chi,\vec{x})}{2(dx)^2a^2}+V(\phi_{\vec{x}},\chi_{\vec{x}})\bigg).
\end{equation}
To solve the Hamiltonian system numerically, we define for time step $h=\delta \tau$ the transformations
\begin{equation}
\Phi_1(h):\bigg(a,p_a,\phi_{\vec{x}},\pi_{\phi,\vec{x}},\chi_{\vec{x}},\pi_{\chi,\vec{x}}\bigg)\rightarrow \bigg(a+\frac{\partial\mathcal{H}_1}{\partial p_a}h,p_a,\phi_{\vec{x}},\pi_{\phi,\vec{x}},\chi_{\vec{x}},\pi_{\chi,\vec{x}}\bigg),
\end{equation}
\begin{equation}
\Phi_2(h):\bigg(a,p_a,\phi_{\vec{x}},\pi_{\phi,\vec{x}},\chi_{\vec{x}},\pi_{\chi,\vec{x}}\bigg)\rightarrow \bigg(a,p_a-\frac{\partial\mathcal{H}_2}{\partial a}h,\phi_{\vec{x}}+\frac{\partial\mathcal{H}_2}{\partial \pi_{\phi,\vec{x}}}h,\pi_{\phi,\vec{x}},\chi_{\vec{x}},\pi_{\chi,\vec{x}}-\frac{\partial\mathcal{H}_2}{\partial \chi_{\vec{x}}}h\bigg),
\end{equation}
\begin{equation}
\Phi_3(h):\bigg(a,p_a,\phi_{\vec{x}},\pi_{\phi,\vec{x}},\chi_{\vec{x}},\pi_{\chi,\vec{x}}\bigg)\rightarrow \bigg(a,p_a-\frac{\partial\mathcal{H}_3}{\partial a}h,\phi_{\vec{x}},\pi_{\phi,\vec{x}},\chi_{\vec{x}}+\frac{\partial\mathcal{H}_3}{\partial \pi_{\chi,\vec{x}}}h,\pi_{\chi,\vec{x}}\bigg)
\end{equation}
and
\begin{equation}
\Phi_4(h):\bigg(a,p_a,\phi_{\vec{x}},\pi_{\phi,\vec{x}},\chi_{\vec{x}},\pi_{\chi,\vec{x}}\bigg)\rightarrow \bigg(a,p_a-\frac{\partial\mathcal{H}_4}{\partial a}h,\phi_{\vec{x}},\pi_{\phi,\vec{x}}-\frac{\partial\mathcal{H}_4}{\partial \phi_{\vec{x}}}h,\chi_{\vec{x}},\pi_{\chi,\vec{x}}-\frac{\partial\mathcal{H}_4}{\partial \chi_{\vec{x}}}h\bigg).
\end{equation}
Then the first order symplectic numerical method (see \cite{EH}) has the form
\begin{equation}
\tilde{\Phi}(h)=\Phi_4(h)\circ\Phi_3(h)\circ\Phi_2(h)\circ\Phi_1(h).
\end{equation}
With that and its adjoint (again, see \cite{EH}), we can create the second order symplectic method for this system, namely
\begin{equation}
\Phi(h)\equiv\tilde{\Phi}^*(h/2)\circ\tilde{\Phi}(h/2)=\Phi_1(h/2)\circ\Phi_2(h/2)\circ\Phi_3(h/2)\circ\Phi_4(h)\circ\Phi_3(h/2)\circ\Phi_2(h/2)\circ\Phi_1(h/2).
\end{equation} 
It can be written as the following set of explicit upgrades:
\begin{equation}
a_{n+1/2}=a_n+\frac{h}{2}\frac{\partial\mathcal{H}_1}{\partial p_a}(p_{a,n})
\end{equation} 
\begin{equation}
\tilde{p}_{a,n+1/2}=p_{a,n}-\frac{h}{2}\frac{\partial\mathcal{H}_2}{\partial a}(a_{n+1/2},\pi_{\phi,n},\chi_n)
\end{equation} 
\begin{equation}
\tilde{\pi}_{\chi,\vec{x},n+1/2}=\pi_{\chi,\vec{x},n}-\frac{h}{2}\frac{\partial\mathcal{H}_2}{\partial \chi_{\vec{x}}}(a_{n+1/2},\pi_{\phi,n},\chi_n)
\end{equation} 
\begin{equation}
\phi_{\vec{x},n+1/2}=\phi_{\vec{x},n}+\frac{h}{2}\frac{\partial\mathcal{H}_2}{\partial \pi_{\phi,\vec{x}}}(a_{n+1/2},\pi_{\phi,n},\chi_n)
\end{equation} 
\begin{equation}
\tilde{\tilde{p}}_{a,n+1/2}=\tilde{p}_{a,n+1/2}-\frac{h}{2}\frac{\partial\mathcal{H}_3}{\partial a}(a_{n+1/2},\tilde{\pi}_{\chi,n+1/2})
\end{equation} 
\begin{equation}
\chi_{\vec{x},n+1/2}=\chi_{\vec{x},n}+\frac{h}{2}\frac{\partial\mathcal{H}_3}{\partial \pi_{\chi,\vec{x}}}(a_{n+1/2},\tilde{\pi}_{\phi,n+1/2})
\end{equation} 
\begin{equation}
\tilde{\tilde{p}}_{a,n+1}=\tilde{\tilde{p}}_{a,n+1/2}-h\frac{\partial\mathcal{H}_4}{\partial a}(a_{n+1/2},\phi_{n+1/2},\chi_{n+1/2})
\end{equation} 
Note that this procedure needs only one array of numbers for every variable and for its associated canonical momenta. These variables are modified consecutively.

\section{Supplementary results}

In order to corroborate our claim that in $\alpha$-attractor models it is the spectator field which is mainly responsible for self-resonance and almost immediate reheating, we compare the results presented in
the main text with a number of alternative simulations run on smaller lattices with $N_\mathrm{lattice}=64$. All the plots presented herein support our main hypothesis.

\subsection{Results for $N_\mathrm{lattice}=64$}
\label{app:b0}

\begin{figure}
\centering
\includegraphics[width=0.495 \textwidth, height=0.29\textwidth]{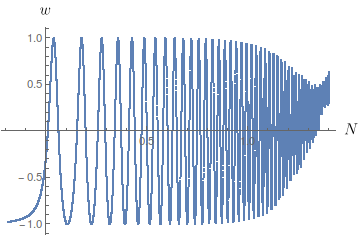}
\includegraphics[width=0.495 \textwidth, height=0.29\textwidth]{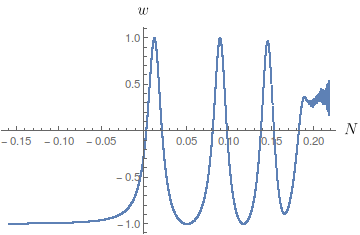} \\
\includegraphics[width=0.495 \textwidth, height=0.29\textwidth]{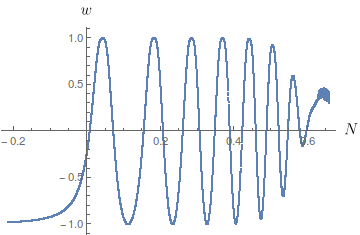}
\includegraphics[width=0.495 \textwidth, height=0.29\textwidth]{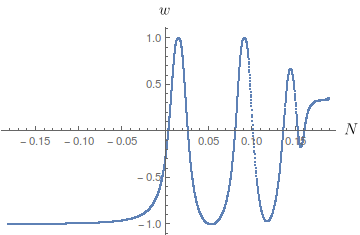}
\caption{\it The evolution of the barotropic parameter $w$ (not averaged) for 
 $n=1,\,\alpha=10^{-3}$ (upper left),  $n=1,\,\alpha=10^{-4}$ (upper right), $n=1.5,\,\alpha=10^{-3}$ (lower left),  $n=1.5,\,\alpha=10^{-4}$ (lower right). 
$N$ is the number of e-folds after the end of inflation.  These results were obtained on a lattice with $N_\mathrm{lattice}=64$.}
\label{wPlot64} 
\end{figure}

\begin{figure}
\centering
\includegraphics[width=0.495 \textwidth, height=0.2\textwidth]{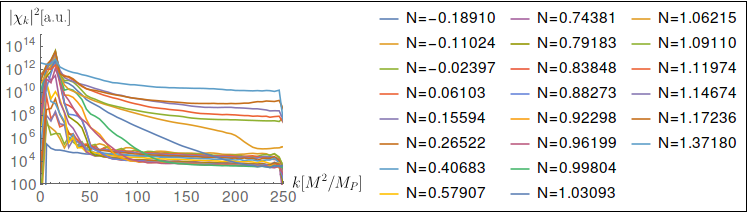} 
\includegraphics[width=0.495\textwidth, height=0.2\textwidth]{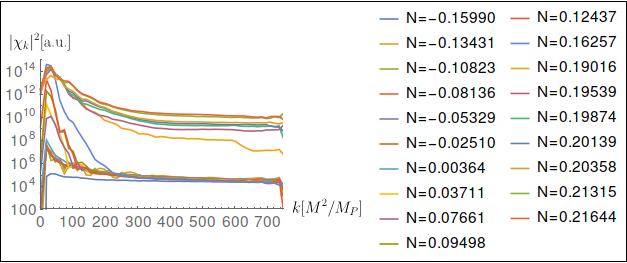} \\
\includegraphics[width=0.495 \textwidth, height=0.2\textwidth]{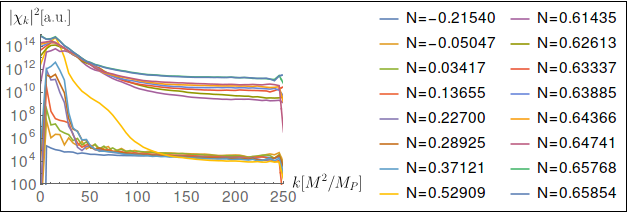} 
\includegraphics[width=0.495\textwidth, height=0.2\textwidth]{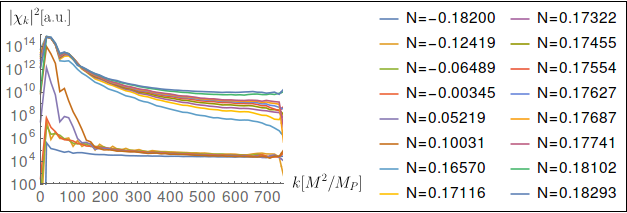} \\
\caption{\it The power spectrum of the spectator perturbations for 
 $n=1,\alpha=10^{-3}$ (upper left),  $n=1,\alpha=10^{-4}$ (upper right), $n=1.5,\alpha=10^{-3}$ (lower left),  $n=1.5,\alpha=10^{-4}$ (lower right)
for different number of e-folds $N$ after the end of inflation. These results were obtained on a lattice with $N_\mathrm{lattice}=64$.} 
\label{PowerSpectrum64}
\end{figure}

\begin{figure}
\centering
\includegraphics[width=0.495 \textwidth, height=0.2\textwidth]{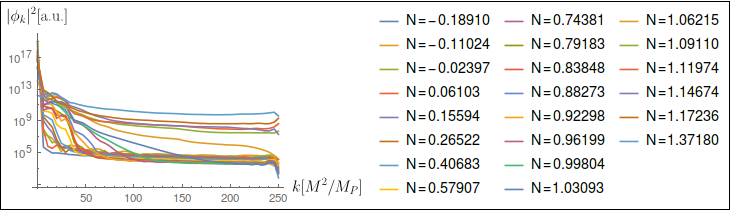} 
\includegraphics[width=0.495\textwidth, height=0.2\textwidth]{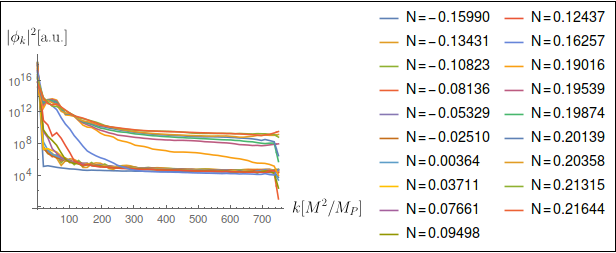} \\
\includegraphics[width=0.495 \textwidth, height=0.2\textwidth]{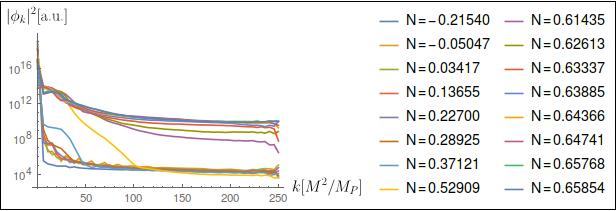} 
\includegraphics[width=0.495\textwidth, height=0.2\textwidth]{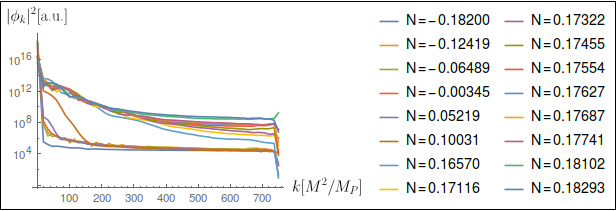} \\
\caption{\it The power spectrum of the inflaton perturbations for 
 $n=1,\,\alpha=10^{-3}$ (upper left),  $n=1,\,\alpha=10^{-4}$ (upper right), $n=1.5,\,\alpha=10^{-3}$ (lower left),  $n=1.5,\,\alpha=10^{-4}$ (lower right)
for different number of e-folds $N$ after the end of inflation. These results were obtained on a lattice with $N_\mathrm{lattice}=64$.} 
\label{PowerSpectrum264}
\end{figure}

\begin{figure}
\centering
\includegraphics[width=0.495 \textwidth, height=0.29\textwidth]{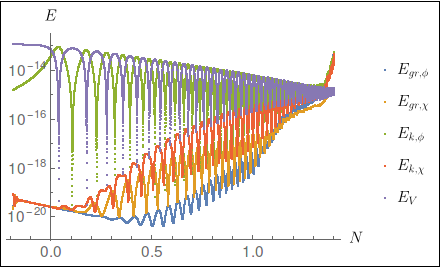}
\includegraphics[width=0.495 \textwidth, height=0.29\textwidth]{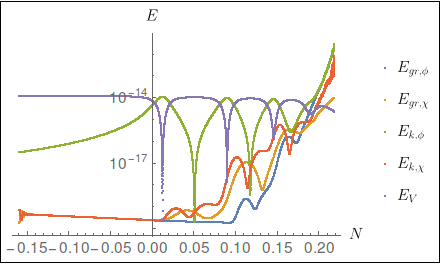} \\
\includegraphics[width=0.495 \textwidth, height=0.29\textwidth]{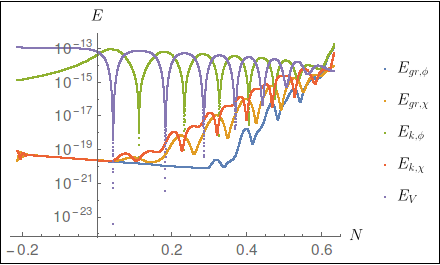}
\includegraphics[width=0.495 \textwidth, height=0.29\textwidth]{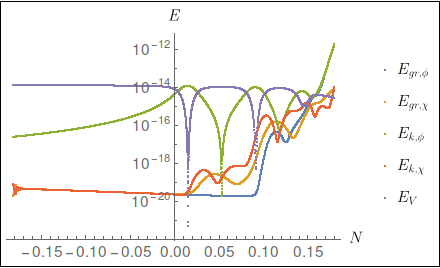}
\caption{\it The evolution of the components of the energy density for 
 $n=1,\,\alpha=10^{-3}$ (upper left),  $n=1,\,\alpha=10^{-4}$ (upper right), $n=1.5,\,\alpha=10^{-3}$ (lower left),  $n=1.5,\,\alpha=10^{-4}$ (lower right). 
$N$ is the number of e-folds after the end of inflation. These results were obtained on a lattice with $N_\mathrm{lattice}=64$.}
\label{fig:en64} 
\end{figure}

The plots presented here show the results obtained according to the procedure described in the main text but on a smaller lattice with $N_\mathrm{lattice}=64$. 
In Figures \ref{wPlot64}, \ref{PowerSpectrum64}, \ref{PowerSpectrum264} and \ref{fig:en64}, we show, respectively, the evolution of the barotropic parameter $w$, the power spectra of the fields $\chi$ and $\phi$, and the contributions to the energy density of the Universe. The results are consistent with those presented in Section \ref{sec:res}, but the simulation for $n=1,\,\alpha=10^{-3}$ was run longer before being stalled by instabilities and the regime $w\to1/3$ is visible.

\subsection{Results for different cutoffs}
\label{app:b12}

\begin{figure}
\centering
\includegraphics[width=0.495 \textwidth, height=0.25\textwidth]{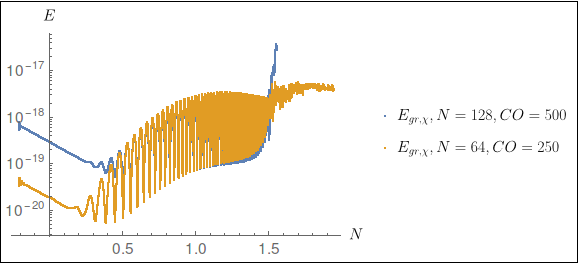} 
\includegraphics[width=0.495\textwidth, height=0.25\textwidth]{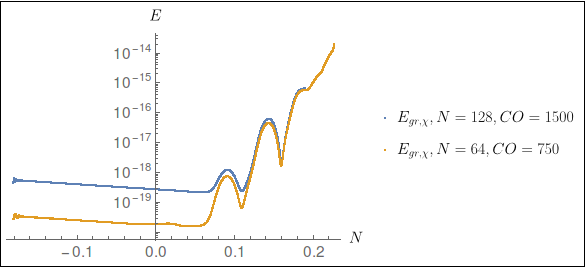} \\
\includegraphics[width=0.495 \textwidth, height=0.25\textwidth]{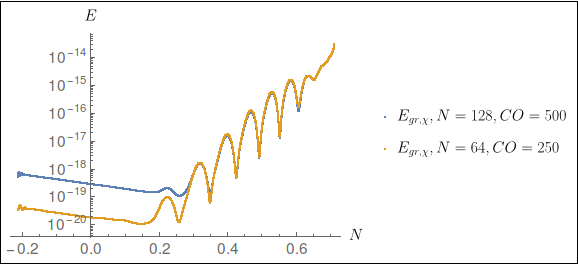} 
\includegraphics[width=0.495\textwidth, height=0.25\textwidth]{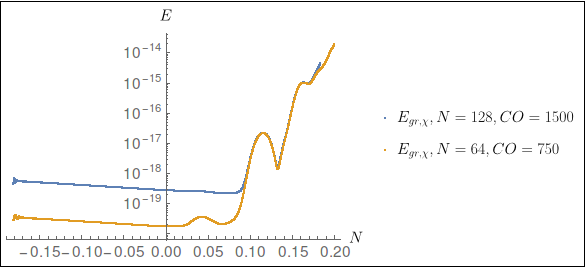} \\
\caption{\it The evolution of gradient energy density of the spectator perturbations for 
 $n=1,\alpha=10^{-3}$ (upper left),  $n=1,\alpha=10^{-4}$ (upper right), $n=1.5,\alpha=10^{-3}$ (lower left),  $n=1.5,\alpha=10^{-4}$ (lower right)
with the number of e-folds $N$ after the end of inflation. Details of the simulations are described in the text.} 
\label{GradEnergyCO}
\end{figure}

In Section \ref{sec:deslat}, we mentioned that our choice of the momentum space cutoff ($k_\mathrm{max}=250M^2/M_P$ and $k_\mathrm{max}=750M^2/M_P$ for $\alpha=10^{-3}$ and $\alpha=10^{-4}$ respectively) is a good trade-off between granularity in Floquet instability regions and factoring in higher frequency modes. We would like to corroborate this statement by comparing simulations with this cutoff on a on a lattice with $N_\mathrm{lattice}=64$ to simulations with a double cutoff ($k_\mathrm{max}=500M^2/M_P$ and $k_\mathrm{max}=1500M^2/M_P$ for $\alpha=10^{-3}$ and $\alpha=10^{-4}$ respectively) on a lattice with a double linear size $N_\mathrm{lattice}=128$. This allows for a direct comparison between the respective simulations, because all the modes present in the simulation on a smaller lattice with $N_\mathrm{lattice}=64$ are also present in the simulation on a bigger lattice with $N_\mathrm{lattice}=128$,
yet the larger simulation involves many modes that are not present in the smaller simulation.

In Figure \ref{GradEnergyCO}, we show the evolution of the gradient energy density of the spectator field in our four benchmark models. Since the perturbations of the spectator field are primarily responsible for the dynamics of the
reheating, this quantity is crucial for determining whether different simulations yield physically identical results. We note that initially the energy density is larger by a factor of 16 for the larger lattices, which is consistent with the fact
that the gradient energy density scales as $k_\mathrm{max}^4$ in the linear regime. However, in a fraction of an efold after the end of inflation, when the nonlinear effects kick in, the calculated gradient energy density depends very weakly on the choice of the cutoff and this agreement is maintained until the numerical instabilities stall the simulation on a larger lattice. Therefore, we can conclude that our cutoff choice has no impact on the physical results.

\end{appendix}

\end{document}